\documentclass[aps,longbibliography,showpacs,twocolumn,superscriptaddress,amsmath,amssymb,verbatim]{revtex4-2}
\usepackage[thinlines]{easytable}
\usepackage{graphicx}
\usepackage{lmodern}
\usepackage{epstopdf}
\usepackage{dcolumn}
\usepackage{bm}
\usepackage{subfigure}
\usepackage{makecell}
\usepackage{amsmath} 
\usepackage[pdftex,colorlinks=true,citecolor=blue,linkcolor=blue,urlcolor=blue,bookmarks=true]{hyperref}

\usepackage[utf8]{inputenc}

\usepackage{xcolor}
\usepackage{tikz}
\usepackage[version=4]{mhchem}

\usepackage{makecell}

\usepackage{color}
\usepackage{textcomp}
\definecolor{red}{rgb}{1,0,0}

\definecolor{blue}{rgb}{0,0,1}

\definecolor{green}{rgb}{0,1,0}

\begin{document}
	\preprint{APS}

\title{Magnetic properties of a quasi-two-dimensional spin-1/2 antiferromagnet Y$_{2}$CuGe$_{4}$O$_{12}$}
\author{J. Khatua}
\affiliation{Department of Physics, Indian Institute of Technology Madras, Chennai 600036, India}
\affiliation{Department of Physics, Sungkyunkwan University, Suwon 16419, Republic of Korea}
\author{Changhyun Koo}
\affiliation{Department of Physics, Sungkyunkwan University, Suwon 16419, Republic of Korea}
\author{Suyoung Kim}
\affiliation{Department of Physics, Simon Fraser University, Burnaby, British Columbia V5A 1S6, Canada}
\author{Eundeok Mun}
\affiliation{Department of Physics, Simon Fraser University, Burnaby, British Columbia V5A 1S6, Canada}
\author{Yugo Oshima}
\affiliation{RIKEN Pioneering Research Institute, Wako, Saitama 351-0198, Japan}
\author{V. K. Sahu}
\affiliation{Department of Physics, Indian Institute of Technology Tirupati, Tirupati 517619, India}
\author{Heung-Sik Kim}
\affiliation{Department of Energy Technology, Korea Institute of Energy Technology, Naju-si 58217, Republic of Korea}
\author{B. Koteswararao}
\affiliation{Department of Physics, Indian Institute of Technology Tirupati, Tirupati 517619, India}
\author{Kwang-Yong Choi}
\email[]{choisky99@skku.edu}
\affiliation{Department of Physics, Sungkyunkwan University, Suwon 16419, Republic of Korea}
\author{P. Khuntia}
\email[]{pkhuntia@iitm.ac.in}
\affiliation{Department of Physics, Indian Institute of Technology Madras, Chennai 600036, India}
\affiliation{Quantum Centre for Diamond and Emergent Materials, Indian Institute of Technology Madras,
	Chennai 600036, India.}

\date{\today}

\begin{abstract}
Competing magnetic interactions and frustration-induced quantum fluctuations in spatially anisotropic low-dimensional magnets often lead to exotic magnetic phenomena, including field-induced phases. Herein, we present the crystal structure, magnetic susceptibility, specific heat, and electron spin resonance (ESR) investigations of polycrystalline Y$_{2}$CuGe$_{4}$O$_{12}$, supported by density functional theory (DFT) calculations, in which the Cu$^{2+}$ ions form a distorted triangular lattice. DFT calculations reveal competing intraplanar ferromagnetic ($J_{1}\simeq0.138$ K and $J_{2}\simeq0.01$ K) and antiferromagnetic ($J_{3}\simeq-3.22$ K) exchange interactions, together with a sub-dominant interplanar antiferromagnetic coupling ($J_{4}\simeq-1.56$ K), consistent with the small Curie--Weiss temperature, $\theta_{\rm CW}=-1.8$ K. Nevertheless, no signature of long-range magnetic ordering is observed down to at least 0.4 K. Instead, the presence of broad maxima in both the magnetic susceptibility and magnetic specific heat indicates the development of short-range spin correlations at low temperatures, as corroborated by the critical ESR line broadening, a characteristic feature of low-dimensional frustrated magnets.  Upon the application of an external magnetic field, the broad maximum in the magnetic specific heat is suppressed, revealing a  competition between the Zeeman and exchange energy scales, which gradually drives the system toward a field-polarized state above the saturation field, $\mu_{0}H_{\rm s} = 2.6$ T. Above this saturation field, the exponential behavior of the magnetic specific heat points to the presence of gapped magnon excitations. Our work identifies Y$_2$CuGe$_4$O$_{12}$ as a rare example of a distorted triangular lattice magnet in which further-neighbor exchange interactions dominate the magnetic behavior, providing a promising platform for investigating frustration-induced quantum phenomena.

\end{abstract}
\maketitle
\section{Introduction}
Quantum magnets composed of spin-1/2 moments are of fundamental interest for exploring correlated quantum phases beyond classical magnetic order~\cite{Balents2010,KHATUA20231,Khuntia2020,Jeon2024,nv1l-dgv9,PhysRevLett.116.107203}. Over the past few decades, a rich spectrum of emergent quantum phenomena has been reported, including  spinons excitations~\cite{PhysRevLett.70.4003,Mourigal2013},  Bose--Einstein condensation of  quasiparticles~\cite{Regg2003}, quantum entanglement~\cite{PhysRevB.103.224434}, and spin-nematic phases in low-dimensional quantum magnets~\cite{PhysRevB.86.014423,10.1063/9.0000401,PhysRevB.90.134401}. In particular, when these systems consist of multiple coupled spin sublattices, the interplay of competing exchange interactions,  spatial anisotropy and reduced dimensionality often harbors exotic excitations, and field-induced phases~\cite{Gen2025,qypv-n4yg}, thus offering a versatile platform for studying interactions among different types of quasiparticles such as magnons and spinons~\cite{PhysRevLett.125.037204,PhysRevLett.87.227201} and their bound states~\cite{PhysRevB.108.L020402,PhysRevB.86.014423}.  Consequently, investigating a broad class of such materials sheds important insights into how dimensional crossover modifies the nature of fractionalized excitations and associated spin dynamics, which are well established in one-dimensional quantum magnets~\cite{JPSJ.76.053705}.\\ In this context, SeCuO$_3$ provides a representative example, where antiferromagnetic dimers and weakly coupled spin chains are linked by weak ferromagnetic interactions, generating frustration and constituting an effective three-dimensional network that stabilizes a half-magnetization plateau alongside magnon, spinon, and triplon excitations~\cite{PhysRevB.86.054405,PhysRevB.103.L020409,PhysRevB.107.054407}. Another notable instance is rouaite, Cu$_2$(OH)$_3$NO$_3$, in which alternating ferromagnetic and antiferromagnetic Cu$^{2+}$ chains are connected by interchain couplings to form a spatially anisotropic triangular lattice~\cite{LINDER19951}. Combined theoretical and single-crystal experimental studies reveal coexisting short-range resonating-valence-bond correlations and weak long-range stripe order in its ground state, together with field-induced phases, including a $2/3$ magnetization plateau and spinon excitations arising from reduced spin dimensionality and competing exchange interactions~\cite{PhysRevB.106.085119,PhysRevB.111.064409}. 
 \\
A different scenario emerges in systems featuring alternating antiferromagnetic chains are found to stabilize conventional collinear magnetic order owing to complex spin lattice with frustrated interplanar interactions, as observed in \ce{Li2CuW2O8}~\cite{PhysRevB.92.094426,lvarezVega2001}, in which second-nearest-neighbor couplings form square motifs, while third-nearest-neighbor diagonal couplings generate an effective triangular lattice.  A related compound, \ce{La2CuGe2O8}, possessing a similar chain--square exchange topology, instead exhibits noncollinear magnetic order at low temperatures~\cite{8zkd78q3,PhysRevB.95.144404}. In this compound, although its network is  formed by first- and second-nearest-neighbor interactions, its third-neighbor couplings generate a distorted triangular lattice.\\While the isotropic triangular lattice provides a platform for frustration driven exotic phases~\cite{PhysRevB.95.014425}, the stability of these phases in the presence of anisotropic hopping arising from unequal bond lengths remains an outstanding issue~\cite{PhysRevB.103.235132}. Furthermore, anisotropic triangular systems offer an important route for investigating the intermediate regime between the Néel order of the square lattice and the 120$^{\circ}$ spiral order of the isotropic triangular lattice~\cite{PhysRevB.103.235132}. Moreover, in such systems, variations  in the relative magnitude and sign of the exchange couplings strongly modify the balance of magnetic interactions, lading to  enhanced frustration and  competition between distinct magnetic ground states. Theoretical studies of anisotropic triangular lattice models have demonstrated that bond dependent exchange anisotropy can substantially alter the magnetic phase diagram and promote competition among collinear, spiral, and quantum-disordered phases~\cite{PhysRevB.86.184421,PhysRevB.88.155139,PhysRevLett.98.077205}.  To this end, the study of diverse triangular lattice systems without anti-site disorder may guide further analytical and numerical efforts toward understanding the proposed frustration induced ground state. \\ 
 \begin{figure*}
 	\centering
 	\includegraphics[width=\textwidth]{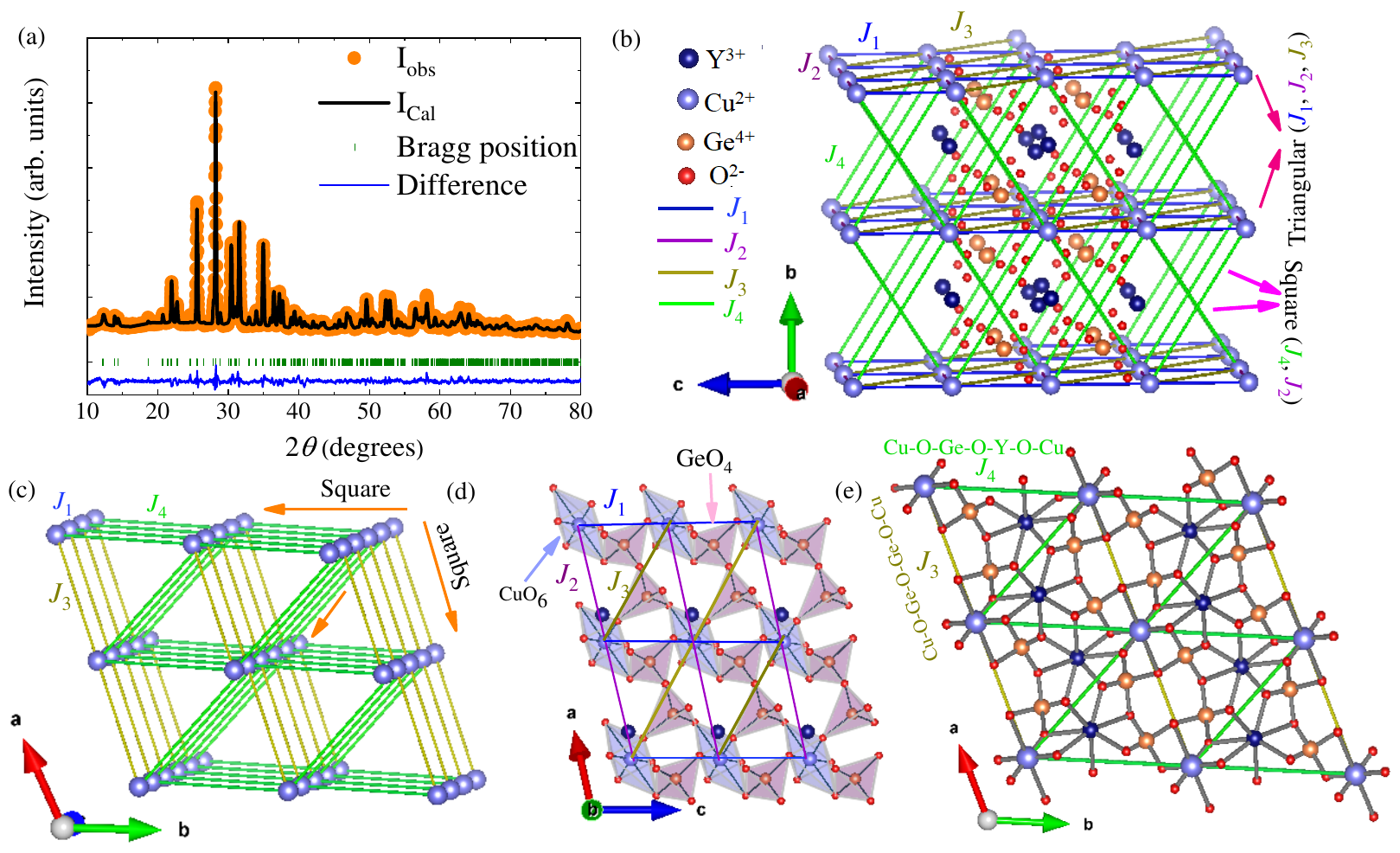}
 	\caption{  (a) Rietveld refinement pattern of the room-temperature powder x-ray diffraction data of Y$_{2}$CuGe$_{4}$O$_{12}$ at room-temperature. The orange circle, black line, olive vertical bars, and blue line denote the experimentally observed points, the result of Rietveld fitting, expected Bragg reflection positions, and the difference between observed and calculated intensities, respectively. (b) Schematic view of the crystal structure of Y$_{2}$CuGe$_{4}$O$_{12}$.
 	The	Cu$^{2+}$ ions form a distorted triangular lattice, which are stacked along the $b$-axis. The distorted triangular network is composed of three inequivalent Cu--Cu exchange pathways: the nearest-neighbor interaction $J_{1}$ corresponding to a Cu--Cu distance of 4.90~\AA\ (blue bonds), $J_{2}$ associated with a Cu--Cu distance of 7.17~\AA\ (purple bonds), and $J_{3}$ corresponding to a Cu--Cu distance of 7.75~\AA\ (dark-yellow bonds). In addition, adjacent triangular layers are connected through the interplanar exchange interaction $J_{4}$ with a Cu--Cu separation of 9.09~\AA\ (green bonds). (c) Three-dimensional square-spin topology with the dominant exchange pathways. (d) Top view of the triangular plane in which distorted CuO$_{6}$ octahedra are formed by nearest-neighbor oxygen ligands of Cu$^{2+}$ ions and are  linked via GeO$_{4}$ tetrahedra. (e) Depicts the exchange pathways associated with the dominant interactions $J_{3}$ and $J_{4}$. }{\label{YCGO1}}.
 \end{figure*}
Herein, we present the crystal structure, magnetic susceptibility, specific heat, and electron spin resonance (ESR) studies of the previously unexplored quasi-two-dimensional frustrated magnet Y$_{2}$CuGe$_{4}$O$_{12}$, in which the Cu$^{2+}$ ions form a distorted triangular lattice within the $ac$ plane comprising three inequivalent intraplanar exchange interactions ($J_{1}\!\sim\!0.138$ K, $J_{2}\!\sim\!0.01$ K, and $J_{3}\!\sim\!-3.22$ K), together with an interplanar coupling ($J_{4}\!\sim\!-1.56$ K). Magnetic susceptibility measurements reveal dominant antiferromagnetic interactions without any signature of long-range magnetic ordering down to at least 0.4 K. The observation of broad maxima in both magnetic susceptibility and specific heat indicates the development of short-range spin correlations at low temperatures. The presence of short-range spin correlations is further supported by the ESR measurements. Furthermore, the hierarchy of spatially anisotropic exchange interactions is consistent with the experimentally observed weak net magnetic energy scale arising from competing interactions. The low saturation field of $\mu_{0}H_{\rm s} = 2.6$ T, required to fully polarize the system, further confirms the coexistence of oppositely signed exchange interactions.
\section{Experimental details}  Polycrystalline samples of Y$_{2}$CuGe$_{4}$O$_{12}$ (hereafter YCGO) were prepared by a conventional solid-state  reaction  method. High-purity starting materials, \ce{Y2O3} (99.997\%), \ce{CuO} (99.995\%), and \ce{GeO2} (99.999\%) from Alfa Aesar, were weighed in stoichiometric proportions and thoroughly ground. To remove residual moisture and carbonate impurities, \ce{Y2O3} was pretreated in air at 900$\,^{\circ}\mathrm{C}$. The homogeneous powder mixture was pressed into pellets and initially heated in an alumina crucible at 700$\,^{\circ}\mathrm{C}$ for 24 h. Several successive grinding and heat-treatment steps at intermediate temperatures were carried out to improve phase homogeneity, followed by a final annealing at 1100$\,^{\circ}\mathrm{C}$ for 48 h, resulting in a single-phase sample.
\\  \begin{table}[b]
	\caption{\label{table}  Structural parameters of Y$_2$CuGe$_4$O$_{12}$ obtained from Rietveld refinement of powder X-ray diffraction data at 300 K. (Space group: $P$ -$1$, $ \alpha$ =  86.79$^{\circ}$, $ \beta$ = 102.71$^{\circ}$, $\gamma$ = 113.86$^{\circ}$), $a$ = 7.171(2) {\AA}, $b$ = 7.938(3) {\AA}, $c$ = 4.899(1) {\AA}
		and $\chi^{2}$ = 3.24, R$_{wp}$ = 4.5 \text{\%}, R$_{p}$ = 3.2 \text{\%}, and R$ _{exp}$ = 4.55\text{\%})}
	\begin{tabular}{c c c c c  c c} 
		\hline \hline
		Atom & Wyckoff position & \textit{x} & \textit{y} &\textit{ z}& Occ. & U$_{\rm iso}$\\
		\hline 
		Y & 2$i$ & 0.233 & 0.455 & 0.041 & 1 & 0.013 \\
		Cu & 1$a$ & 0.000 & 0.000 & 0.00 & 1 & 0.010\\
		Ge$_{1}$ & 2$i$ & 0.380& 0.201 & $-$0.420 & 1 & 0.009 \\
		Ge$_{2}$ & 2$i$ & 0.169& 0.780 & $-$0.450 & 1 & 0.012 \\
		O$_{1}$ & 2$i$ & 0.177& 0.159 & $-$0.807 & 1 & 0.011 \\
		O$_{2}$ & 2$i$ & $-$0.029& 0.811 & $-$0.775 & 1 & 0.010 \\
		O$_{3}$ & 2$i$ & 0.312& $-$.010 & $-$ 0.193 & 1 & 0.008\\
		O$_{4}$ & 2$i$ & 0.358& 0.735 & 0.427 & 1 & 0.015\\
		O$_{5}$ & 2$i$ & 0.436& 0.367 & $-$0.230 & 1 & 0.010 \\
		O$_{6}$ & 2$i$ & 0.084& 0.628 & $-$0.259 & 1 & 0.010 \\	
		\hline
	\end{tabular}
\end{table}
Room-temperature powder x-ray diffraction (XRD) measurements were carried out using a PANalytical X$'$Pert PRO diffractometer equipped with Cu K$_{\alpha}$ radiation ($\lambda = 1.54$~\AA).\\ Magnetization measurements were performed using  a superconducting quantum
interference device vibrating sample magnetometer (SQUID-VSM, Quantum Design) in the temperature range 2 K $\leq$ \textit{T} $\leq$ 300 K and in magnetic fields up to 7 T.  Furthermore, low-temperature  magnetic susceptibility and magnetization isotherms were measured down to 0.4 K using the $^3$He option of the Quantum Design SQUID magnetometer.\\
 Specific heat measurements were conducted  using a Quantum Design, physical properties measurement system (PPMS) in the temperature range 2 K $\leq$ \textit{T} $\leq$ 250 K and in magnetic fields up to 9 T. For measurements in the temperature range  0.4 K $\leq$ \textit{T} $\leq$ 4 K, the $^{3}$He option of the Quantum Design Dynacool PPMS was used.\\  ESR measurements were performed with a commercial X-band ESR spectrometer (JES-REX3X, JEOL) at a fixed frequency $f = 9.12$ GHz in RIKEN. A polycrystal of YCGO compound was loaded on the tip of a non-magnetic quartz stick using silicon grease. The magnetic field was swept from 245 mT to 345 mT. A continuous-flow Helium cryostat (ESR900, Oxford Instruments) was employed to control temperatures in the temperature range of 3.6 K $\leq$ $T$ $\leq$ 300 K during the ESR measurements. \\
Density functional theory (DFT) calculations were performed using the OpenMX code within the generalized gradient approximation (GGA)+$U$ framework employing the Perdew–Burke–Ernzerhof (PBE) exchange-correlation functional~\cite{PhysRevB.67.155108}. Strong electronic correlations on the Cu sites were treated using different Hubbard $U$ values  to examine the robustness of the calculated exchange couplings. The calculations were carried out with an energy cutoff of 300 Ry and self-consistent field (SCF) convergence criterion of 1.0 $\times$ $10^{-9}$ Hartree, using a $2\times1\times2$ supercell. For the determination of exchange interactions, the  converged electronic structure obtained from OpenMX was subsequently employed as input for the $J_{\rm X}$ code~\cite{YOON2020106927}, which evaluates the magnetic exchange parameters $J_{ij}^{\rm GGA}$ between localized spins within the Green’s-function-based Liechtenstein formalism.
\section{RESULTS AND DISCUSSION}
\subsection{Rietveld refinement  and crystal structure of Y$_{2}$CuGe$_{4}$O$_{12}$ } 
To ensure the phase purity and extract the crystallographic parameters of polycrystalline YCGO, Rietveld refinement of the powder XRD data was carried out using the GSAS software package~\cite{Toby:hw0089}. The refinement was initiated using structural parameters reported in Refs.~\cite{CASCALES2002379,Cascales2000}. The calculated diffraction profile based on the triclinic $P\bar{1}$ space group reproduces the experimental diffraction pattern very well, as shown in Fig.~\ref{YCGO1}(a). No impurity reflections were detected, confirming the formation of single-phase of YCGO. The refined lattice constants together with the goodness factors are summarized in Table~\ref{table}. These results indicate  the absence of any detectable atomic site disorder in YCGO. In addition, considering the substantial differences in ionic radii between Cu$^{2+}$ ($r_{\rm Cu^{2+}} = 0.73$~\AA), Y$^{3+}$ ($r_{\rm Y^{3+}} = 0.90$~\AA), and Ge$^{4+}$ ($r_{\rm Ge^{4+}} = 0.53$~\AA), anti-site mixing between the magnetic Cu ions and the nonmagnetic Y/Ge ions is expected to be highly unfavorable.\\ \\
The magnetic spin topology of YCGO can be described as a distorted triangular network of Cu$^{2+}$ ions within the $ac$-plane~(Fig.~\ref{YCGO1}(b)). The nearest-neighbor Cu$^{2+}$ ions are connected through Cu--Cu bonds of length 4.90~\AA\ (blue bonds), corresponding to the exchange interaction $J_{1}$. The second-nearest-neighbor Cu$^{2+}$ ions, separated by 7.17~\AA\ (purple bonds), are coupled via the interaction $J_{2}$, while the third-nearest-neighbor Cu$^{2+}$ ions, with a bond length of 7.75~\AA\ (dark-yellow bonds), are connected through the interaction $J_{3}$. The combined effect of these inequivalent exchange pathways generates a distorted triangular arrangement of magnetic interactions within the $ac$-plane, providing the basis for geometric frustration in this system. The resulting triangular planes are further coupled by an interplanar exchange interaction $J_{4}$ over a distance of 9.09~{\AA} (Fig.~\ref{YCGO1}(b)). In the limit where $J_{2}$ is negligible, such an anisotropic triangular lattice can be treated as alternating zigzag spin chains.  Figure~\ref{YCGO1}(c) presents the spin topology constructed from the $J_{1}$, $J_{3}$, and $J_{4}$ exchange interactions, excluding the comparatively weak $J_{2}$ interaction identified from the DFT calculations (see Sec.~\ref{mag}). While several additional interplanar Cu–Cu pairs indeed exist between 7.75 and 9.09 {\AA}, our DFT calculations show that their corresponding exchange interactions are negligible.  \\ In YCGO, the Cu$^{2+}$ ions form distorted CuO$_{6}$ octahedra coordinated by O$^{2-}$ ions, which are arranged in chains extending along the crystallographic $c$-axis (see Fig.~\ref{YCGO1}(d)).  The connectivity between the CuO$_{6}$ octahedra and GeO$_{4}$ tetrahedra gives rise to Cu--O--Ge--O--Cu superexchange pathways along both the $J_{1}$ and $J_{2}$ routes. The different Cu--Cu bond lengths are color-coded for clarity (Fig.~\ref{YCGO1}(d)).
\\
The variation in bond lengths and bond angles along the Cu--O--Ge--O--Cu exchange pathways is expected to yield different strengths of the $J_{1}$ and $J_{2}$ exchange interactions. In addition to the first- and second-nearest-neighbor exchange paths, the longer Cu--O--Ge--O--Ge--O--Cu exchange pathway ($J_{3}$; Fig.~\ref{YCGO1}(e)), which connects the diagonal sites of the square lattice,  provides an additional exchange channel associated with the distorted triangular magnetic network in YCGO. Similarly, the $J_{4}$ exchange interaction is expected to be mediated through the extended Cu--O--Ge--O--Y--O--Cu superexchange pathway~(Fig.~\ref{YCGO1}(e)).
\\ In general, YCGO is expected to exhibit relatively weak exchange interactions, similar to those in La$_2$CuGe$_2$O$_8$~\cite{8zkd78q3}, as both compounds are linked by Cu--O--Ge--O--Cu superexchange pathways. This originates from the filled $3d^{10}$ electronic configuration of Ge$^{4+}$, whose low-lying, tightly bound $d$ states weakly hybridize with O $2p$ orbitals, thereby suppressing the superexchange interaction. In contrast, the empty and spatially extended $5d$ orbitals of W$^{6+}$ strongly hybridize with O $2p$ states, giving rise to much stronger Cu--O--W--O--Cu superexchange interactions, as observed in \ce{Sr2CuWO6}~\cite{PhysRevMaterials.8.094404,PhysRevB.89.134419}.
 \begin{figure*}
	\centering
	\includegraphics[width=\textwidth]{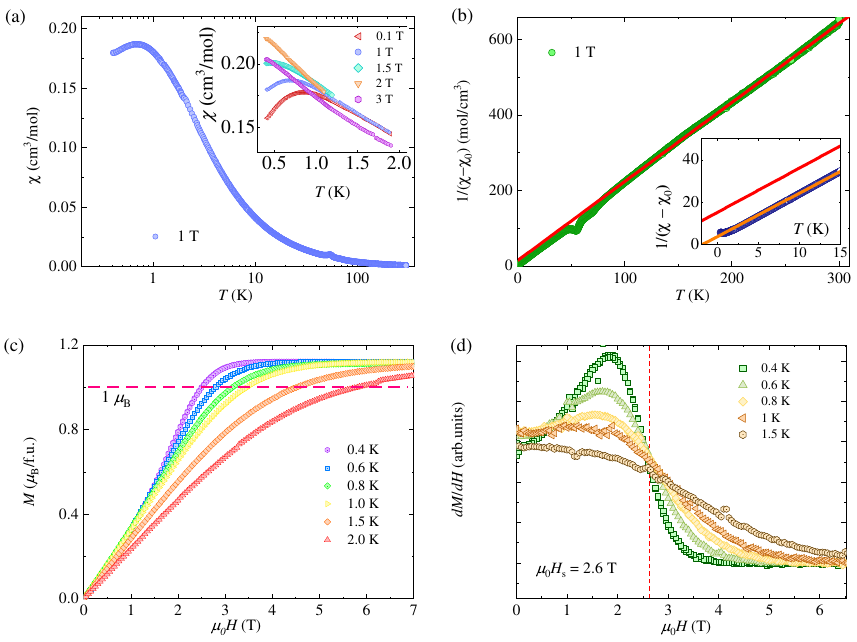}
	\caption{(a) Temperature dependence of magnetic susceptibility, $\chi(T)$, of Y$_{2}$CuGe$_{4}$O$_{12}$ in a magnetic field of $\mu_{0}H = 1$ T. A weak  hump around 50 K is attributed to trapped oxygen in the Teflon-wrapped pellet. The inset depicts $\chi(T)$ in several magnetic fields below 2 K. (b) Temperature dependence of the inverse magnetic susceptibility after subtraction of the temperature-independent $\chi_{0}$ contribution (see text). The red line represents the Curie--Weiss fit to the high-temperature inverse susceptibility data. The bottom inset shows the Curie--Weiss fit in the low-temperature region. (c) Isothermal magnetization as a function of external magnetic field at several temperatures. The horizontal dashed line corresponds to 1 $\mu_{\rm B}$. (d) Field derivative of the isothermal magnetization at several temperatures. The dashed vertical line marks the saturation field, $\mu_{0}H_{\rm s} = 2.6$ T.  }{\label{YCGO2}}.
\end{figure*}
\begin{figure*}[t]
	\centering
	\includegraphics[width=\textwidth]{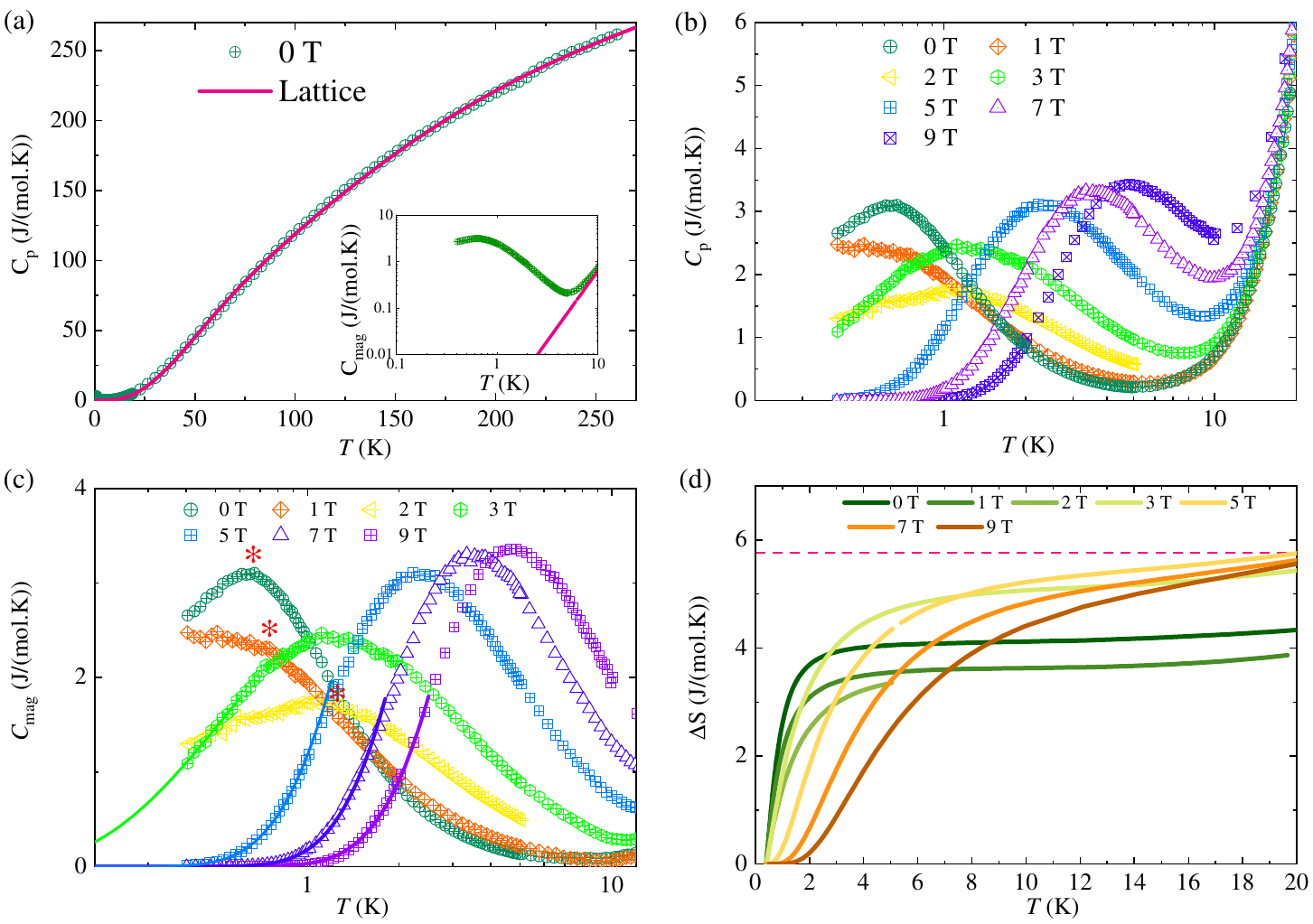}
	\caption{(a) Temperature dependence of the specific heat, $C_{\rm p}(T)$, of Y$_{2}$CuGe$_{4}$O$_{12}$  down to 0.4 K in zero magnetic field. The solid line represents the fit using the Debye--Einstein model for the lattice contribution to the specific heat. The bottom inset shows a zoomed view of the low-temperature \textit{C}$_{\rm p}(T)$ data.
		(b) Temperature dependence of $C_{\rm p}(T)$  under various magnetic fields, plotted on a semi-logarithmic scale.
		(c) Temperature dependence of the magnetic specific heat, $C_{\rm mag}(T)$, in different magnetic fields. The solid lines represent fits using an exponential function accounting for gapped magnon excitations for magnetic fields above the saturation field, $\mu_{0}H_{\rm s} = 2.6$ T.
		(d) Temperature dependence of the magnetic entropy change in various magnetic fields. 
	}{\label{YCGO3}}.
\end{figure*} 
\subsection{Magnetic susceptibility}\label{mag}
Figure~\ref{YCGO2}(a) shows the temperature dependence of the magnetic susceptibility, $\chi(T)$, measured in a magnetic field of $\mu_{0}H = 1$ T. Upon lowering the temperature, $\chi(T)$ increases gradually and rises sharply below 10 K, signaling the onset of magnetic correlations. A broad maximum at $T_{\rm max}=0.70$ K suggests the development of short-range spin correlations, followed by a decrease of $\chi(T)$ at lower temperatures, typical of low-dimensional and frustrated antiferromagnetic systems~\cite{PhysRev.135.A640,PhysRevB.73.134414}.   The absence  of an anomaly indicates that the Cu$^{2+}$ moments remain disordered down to at least 0.4 K. A weak hump around 50 K likely arises from trapped oxygen in the teflon-wrapped pressed pellet sample and is therefore not intrinsic to the material~\cite{PhysRevB.109.184432}. \\ To examine the field dependence of the broad maximum, low-temperature $\chi(T)$ measurements below 2 K were performed in magnetic fields up to 3 T, as shown in the inset of Fig.~\ref{YCGO2}(a). At $\mu_{0}H$ = 0.1 T, the broad maximum appears at $T_{\rm max}=0.86$ K and shifts to lower temperatures with enhanced $\chi(T)$ upon increasing field. However, at 3 T, $\chi(T)$ becomes suppressed, suggesting the possible presence of a critical field near 3 T. The field-induced shift and enhancement of the broad maximum indicate that the Zeeman energy competes with the exchange interaction $J$, thus weakening the antiferromagnetic correlations and driving the system toward a field-polarized state above the saturation field $\mu_{0}H_{s}$.\\
The temperature dependence of the inverse magnetic susceptibility (1/$\chi(T)$) of YCGO, obtained after subtracting the diamagnetic contribution \cite{Bain2008} $\chi_{0}$ = $-$ 3.889 $\times 10^{-4}$ cm$^{3}$/mol, is presented in Fig.~\ref{YCGO2}(b). The linear high-temperature region above 100 K was fitted using a Curie--Weiss (CW) law, $\chi (T) $ $-\chi_{0}$ = $C/(T-\theta_{\rm CW})$ (red solid line), yielding the CW temperature $\theta_{\rm CW}^{\rm HT} = -7.23$ K and the Curie-constant $C = 0.477$ cm$^{3}$K/mol. The estimated effective magnetic moment, $\mu_{\rm eff}^{\rm exp} = \sqrt{8C} = 1.95~\mu_{\rm B}$, is slightly higher than the theoretical spin-only value ($\mu_{\rm eff}^{\rm the} = 1.73~\mu_{\rm B}$) expected for $S = 1/2$ Cu$^{2+}$ moments. The corresponding Landé $g$-factor, obtained using $\mu_{\rm eff}^{\rm exp} = g\sqrt{S(S+1)} \ \ \mu_{\rm B}$ for $S=1/2$, is $g \approx 2.25$. The negative CW temperature indicates the presence of dominant antiferromagnetic interactions in YCGO. As shown in the inset of Fig.~\ref{YCGO2}(b), the low-temperature inverse susceptibility below 40 K follows an additional CW behavior with $\theta_{\rm CW}^{\rm LT} = -1.80$ K, which is smaller than $\theta_{\rm CW}^{\rm HT}$, implying a renormalization of the exchange interactions due to the onset of  possibly sub-dominant ferromagnetic interactions.\\ \begin{table}[ht]
	\centering
	\caption{Exchange interactions obtained for different values of the Hubbard $U$ parameter for Y$_{2}$CuGe$_{4}$O$_{12}$.}
	\begin{tabular}{ccccc}
		\hline
		$U$ (eV) & 2 & 4 & 6 & 8 \\
		\hline
		{$J_{1}$ (4.90 \AA)} 
		& 0.243 K & 0.212 K & 0.172 K & 0.138 K \\
		
		{$J_{2}$ (7.17 \AA)} 
		& 0.033 K & 0.023 K & 0.015 K & 0.010 K \\
		
	{$J_{3}$ (7.75 \AA)} 
		& $-$5.123 K & $-$4.66 K & $-$3.944 K & $-$3.224 K \\
		
		{$J_{4}$ (9.09 \AA)} 
		& $-$2.35 K & $-$2.17 K & $-$1.866 K & $-$1.568 K \\
		\hline
	\end{tabular}
	\label{table:exchange}
\end{table}
To further understand the dominant nature of exchange couplings, we computed the leading exchange interactions in YCGO using DFT calculations, as summarized in Table~\ref{table:exchange}. The results indicate that the nearest-neighbor interaction $J_{1}$ is ferromagnetic, while the second-nearest-neighbor interaction $J_{2}$ is  weakly ferromagnetic. In contrast, the third-nearest-neighbor interaction $J_{3}$ is the dominant antiferromagnetic coupling, accompanied by an additional antiferromagnetic interplanar interaction $J_{4}$. Thus, YCGO exhibits a complex mixture of ferro- and antiferromagnetic interactions with the hierarchy
$|J_{3}| > |J_{4}| > |J_{1}| > |J_{2}|.$ This behavior is broadly consistent with the expectation that the Ge$^{4+}$ ions mediate weaker superexchange interactions than W$^{6+}$ ions, leading to a reduced magnetic energy scale and enhanced competition between exchange interactions. 
Using these exchange parameters, the CW temperature was estimated following
$\theta_{\rm CW} = \frac{S(S+1)}{3k_{\rm B}} \sum_{i} z_{i}J_{i}$.
For $U = 8$ eV and coordination number $z_i = 2$, the calculated value is $\theta_{\rm CW} \approx -2.32$ K, which is slightly larger than the value obtained from the low-temperature CW fit.\\
The isotherm magnetization, $M(H)$, as a function of magnetic field is shown in Fig.~\ref{YCGO2}(c) at several temperatures. The linear behaviour of $M(H)$ at low-fields is consistent with dominant antiferomagnetic interactions. The experimentally observed saturation magnetization reaches $M_{\rm sat}=1.12~\mu_{\rm B}$/f.u., slightly exceeding the theoretically expected value of $1~\mu_{\rm B}$/f.u. for $g=2$, as indicated by the dashed horizontal line (Fig.~\ref{YCGO2}(c)). This enhancement can be attributed to the experimentally obtained Landé $g$-factor of $g \approx 2.25$, yielding $M_{\rm sat} = gS \approx 1.125~\mu_{\rm B}$.  The field derivative of the isothermal magnetization, $dM/dH$ as a function of magnetic field is shown in Fig.~\ref{YCGO2}(d). The saturation field $\mu_{0}H_{\rm s}=2.6$~T, separating the low-field antiferromagnetic regime from the high-field polarized state, is indicated by the dashed vertical line. A pronounced maximum is observed below $\mu_{0}H_{\rm s}=2.6$~T, which broadens and gradually weakens with increasing temperature due to thermal fluctuations. A similar feature has also been observed in the distorted triangular lattice antiferromagnet La$_{2}$CuGe$_{2}$O$_{8}$, which exhibits $\theta_{\rm CW} = -3.74$ K, $T_{\rm N} = 1.19$ K, and $\mu_{0}H_{\rm s} = 3.87$ T~\cite{8zkd78q3}.
\subsection{Specific heat}
\begin{figure}[b]
	\centering
	\includegraphics[width=0.5\textwidth]{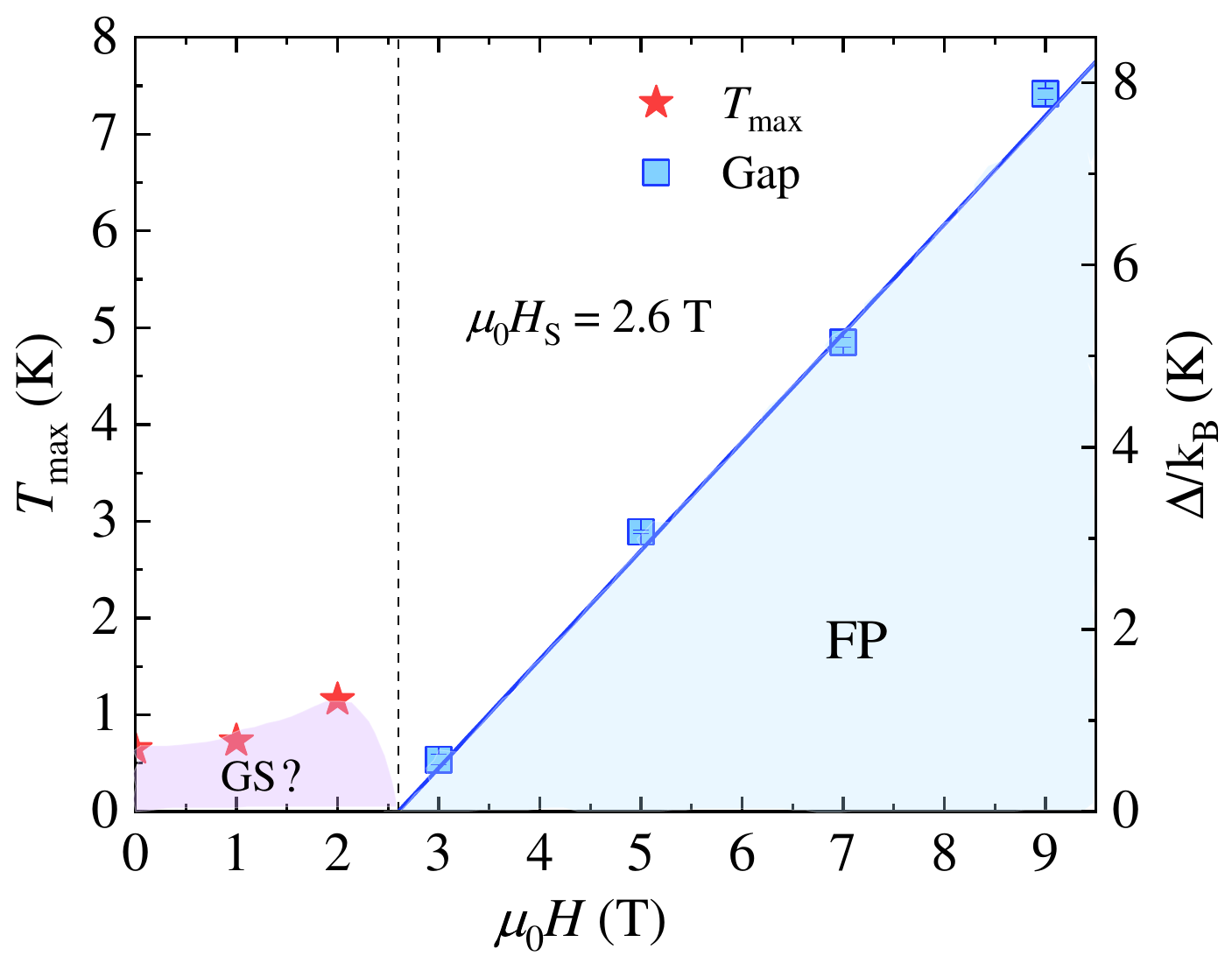}
	\caption{Evolution of $T_{\rm max}$ and the magnon gap $\Delta/k_{\rm B}$ as a function of magnetic field. Red stars indicate the position of the broad maxima ($T_{\rm max}$) in $C_{\rm mag}$, below which the magnetic ground state (GS) remains unclear. Blue squares represent the field-induced magnon gap $\Delta/k_{\rm B}$ in the polarized regime ($H > H_{s}$), showing a linear behavior that vanishes at the saturation field $\mu_{0}H_{s} = 2.6$~T.
	}{\label{YCGO4}}.
\end{figure}

To further investigate the magnetic properties and field-induced behavior of the local moments, specific heat measurements were performed down to 0.4 K in several magnetic fields. The temperature dependence of the specific heat, $C_{\rm p}(T)$, measured in zero field and under several  magnetic fields is shown in Fig.~\ref{YCGO3}(a) and Fig.~\ref{YCGO3}(b), respectively. The absence of any $\lambda$-like anomaly indicates the absence of long-range magnetic order  down to at least 0.4 K, consistent with the $\chi(T)$ data. A broad maximum is observed around $T_{\rm max} = 0.64$ K, which is close to that observed in the $\chi(T)$ data, reflecting the development of short-range spin correlations. \\  In order to estimate the magnetic contribution associated with the Cu$^{2+}$ moments, the magnetic specific heat was obtained using $C_{\rm mag}(T)=C_{\rm p}(T)-C_{\rm latt}(T)$, where $C_{\rm latt}(T)$ represents the estimated phonon contribution. In the absence of a suitable non-magnetic analog, the lattice specific heat was approximated using one Debye and three Einstein terms,
\begin{equation*}
C_{\rm latt}(T)=C_{D}\left[9R\left(\frac{T}{\theta_{D}}\right)^{3}\int_{0}^{\theta_{D}/T}\frac{x^{4}e^{x}}{(e^{x}-1)^{2}}dx\right]
\end{equation*}
\begin{equation}
+\sum_{i=1}^{3} C_{E_i}\left[R\left(\frac{\theta_{E_i}}{T}\right)^{2}
\frac{\exp(\theta_{E_i}/T)}
{\left[\exp(\theta_{E_i}/T)-1\right]^{2}}\right],
\label{eqn:debye}
\end{equation}
where $\theta_D$ and $\theta_{E_i}$ denote the Debye and Einstein temperatures, respectively. The solid red line in Fig.~\ref{YCGO3}(a) represents the fit to the experimental data, yielding $\theta_D = 184$ K, $\theta_{E_1} = 262$ K, $\theta_{E_2} = 502$ K, and $\theta_{E_3} = 1236$ K. While fitting the data, the coefficients were fixed as $C_D = 3$, $C_{E_1} = 12$, $C_{E_2} = 18$, and $C_{E_3} = 24$, corresponding to the three acoustic phonon modes and the remaining $(3n-3)$ optical modes, where $n$ is the total number of atoms in YCGO.
The $C_{p}(T)$ data deviate from the lattice fit below 6 K, suggesting the onset of spin correlations, consistent with the small characteristic exchange energy scale, $\theta_{\rm CW}$.  \\
The temperature dependence of the magnetic specific heat, $C_{\rm mag}(T)$, derived after subtracting the phonon contribution, is shown in Fig.~\ref{YCGO3}(c). Most intriguingly, with increasing magnetic field, the broad maximum in $C_{\rm mag}(T)$ becomes slightly narrower and shifts to higher temperature.  In contrast, $\chi(T)$ shows a pronounced suppression of the broad maximum on approaching a saturation field.  Such a contrast with $\chi(T)$ arises because magnetic susceptibility probes the field response of the spins that remain unquenched by the magnetic field, whereas specific heat is sensitive to the magnetic excitation spectrum and the associated entropy. \\
At 3 T, just above the saturation field $\mu_{0}H_{\rm s}=2.6$ T, the system enters a field-polarized state. In this regime, low-energy spin fluctuations are gapped out, requiring finite energy to overcome the excitation gap. The enhancement of $C_{\rm mag}(T)$ at 3 T compared with 2 T may indicate the accumulation of low-energy magnon states near the saturation field, where the excitation gap is small and thermal population of spin excitations is maximized. With further increasing field, the broad maximum shifts systematically to higher temperature, consistent with the increase of the magnon gap with magnetic field. For the fields $\mu_{0}H>$ $\mu_{0}H_{s}$, at low temperatures, the magnetic specific heat $C_{\rm mag}(T)$ is well described by an exponential form,
$
C_{\rm mag}(T) \sim \exp\left(-\frac{\Delta}{k_{\rm B}T}\right),
$
where $\Delta$ is the magnon gap and $k_{\rm B}$ represents the Boltzmann constant.  The magnetic entropy change, $\Delta S$, was calculated by integrating $C_{\rm mag}(T)/T$ over the temperature range from 0.4 K to 20 K, as shown in Fig.~\ref{YCGO4}(d). While $\Delta S$ approaches the theoretical $R\ln2$ value expected for an $S = 1/2$ system at fields $\mu_{0}H \geq 3$ T, it remains lower for $\mu_{0}H < 3$ T. This missing entropy indicates that a substantial part of the magnetic specific heat lies below 0.4 K in the low-field region and/or reflects an overestimation of the lattice contribution.\\
 Figure~\ref{YCGO4} summarizes the field-temperature phase diagram of YCGO. The star symbols trace the broad maximum in $C_{\rm mag}(T)$ below $\mu_{0}H_{\rm s}$, while the full squares represent the field dependence of the magnon gap in the field-polarized (FP) phase. While the precise nature of the ground state below $\mu_{0}H_{s}$ remains an open question for future investigation, the extracted gap $\Delta$ above saturation exhibits a clear linear field dependence. A linear fit using
 $
 \Delta = g\mu_{B}(\mu_{0}H - \mu_{0}H_{s}),
$
 yields a slope of 1.193 K/T, corresponding to a $g$-factor of approximately 1.78~\cite{nv1l-dgv9}.
\begin{figure}[b]
	\centering
	\includegraphics[width=0.5\textwidth]{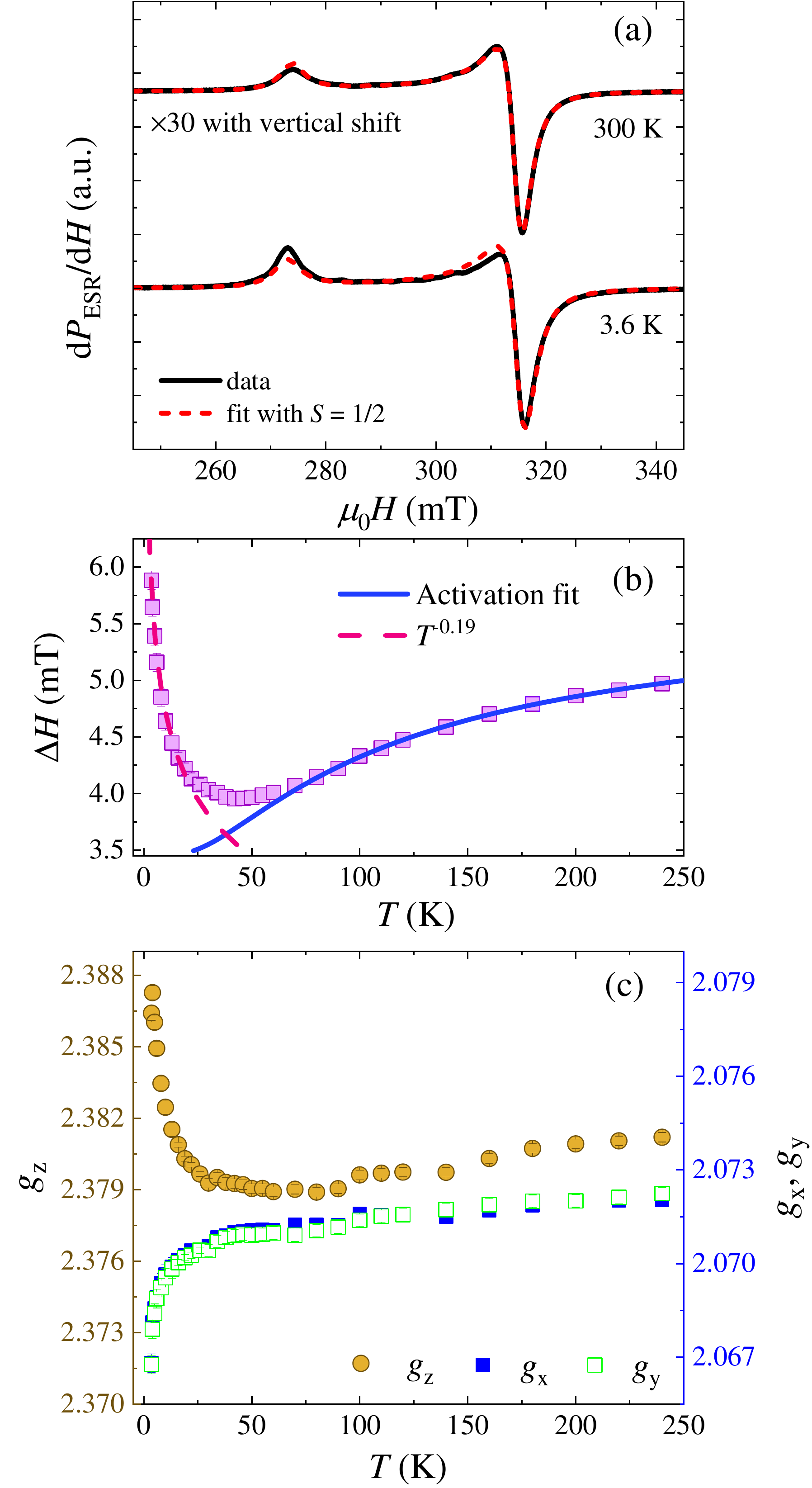}
	\caption{(a) ESR spectra measured at $T = 300$ K and 3.6 K for $f = 9.12$ GHz (Black line). The red dashed line represents a fit to the data. See text for the detail explanation for the fit. (b), (c) Temperature dependence of linewidth and $g$-factor obtained from the fits to the spectra.  
	}{\label{ESR}}.
\end{figure}
\section{Electron spin resonance}
To investigate spin correlations and dynamics in YCGO, ESR measurements were performed over a wide temperature range, 3.6 K to 300 K. The ESR spectrum at 3.6 K exhibits typical powder features with axial $g$-anisotropy of Cu$^{2+}$ ions, consisting of a weak feature around 273 mT and a larger asymmetric absorption line around 314 mT (Fig.~\ref{ESR}(a)). Notably, these powder spectra remain observable throughout the entire temperature range, while both the resonance positions and linewidth exhibit pronounced temperature dependence.\\
To elucidate the temperature-dependent behavior, the spectra were analyzed using the powder ESR fitting program provided by EasySpin \cite{STOLL200642}. A representative fit to the ESR spectra at 3.6 K and 300 K is shown in Fig.~\ref{ESR}(a). In the fitting procedure assuming $S$ = 1/2, the principle-axis $g$-factors and the averaged linewidth of the resonance were treated as fitting parameters. Although the extracted linewidth represents an average value over all crystallographic directions and therefore does not resolve the directional evolution of spin correlations, it provides an approximate estimate of the temperature dependence of the linewidth.
\\
As shown in Fig.~\ref{ESR}(b), the linewidth decreases from approximately 5 mT at 240 K to 3.9 mT at 42 K upon cooling. This behavior deviates from the nearly temperature-independent linewidth typically observed in the high-temperature exchange-narrowing regime, where spin correlations are negligible and the (modified) Kubo--Tomita limit may be approached \cite{doi:10.1143/JPSJ.9.888,PhysRevB.65.134410,PhysRevLett.87.127207}. The observed decrease in linewidth is therefore likely associated with additional relaxation process such as a spin--phonon relaxation process \cite{HUBER1975723,PhysRevB.67.224418,PhysRevLett.101.147601,PhysRevB.94.104408}. The high-temperature linewidth is well described by (blue line in Fig.~\ref{ESR}(b))
$
\Delta H(T)=\Delta H_{0}+A\exp\left(-\frac{\Delta}{k_{B}T}\right),
$ where $\Delta H_{0}$ represents the limiting linewidth at the crossover temperature, while the exponential term reflects a thermally activated spin-lattice relaxation process characterized by an activation gap $\Delta/k_{\rm B}$ = 96 K~\cite{PhysRevB.67.224418}. It is worth noting that the high-temperature linewidth can alternatively be described within the framework of symmetric anisotropic exchange interactions. However, such an analysis did not yield physically reasonable fit parameters, with the extracted energy scales being significantly larger than the dominant exchange interactions estimated  from DFT and thermodynamic experiments in  YCGO~\cite{PhysRevB.95.054405}.\\    Upon further cooling below 42 K, the linewidth increases continuously following a power-law dependence, $\Delta H \propto T^{-0.19}$, commonly observed in frustrated magnetic systems~\cite{PhysRevB.109.184432} and indicative of the development of short-range spin correlations. Notably, this temperature closely coincides with the onset of the deviation of the magnetic susceptibility from the high-temperature CW fit ($T \simeq 40$ K), providing complementary evidence for the growth of spin correlations~(see Fig.~\ref{YCGO2}(b)). Such behavior is typical of frustrated magnetic systems, where significant short-range correlations develop at temperatures well above the characteristic exchange energy scale defined the CW temperature $|\theta_{\rm CW}|$. 
\\
The temperature dependence of the $g$-factors provides more direct evidence for the evolution of the internal magnetic field and its anisotropy (Fig.~\ref{ESR}(c)). Below approximately 40 K, the $g$-factor along the $z$-axis increases continuously, whereas the $g$-factors along the $x$- and $y$-axes decrease upon cooling. This behavior suggests that the internal field along the $z$-axis is weaker than those along the $x$- and $y$-axes at low temperatures, where intraplanar correlation is dominant.  Above 40 K, the $g$-factors for all directions increase gradually with increasing temperature, which may reflect the influence of an additional relaxation mechanism active in the high-temperature regime.
\section{Conclusion}

In conclusion, we have synthesized and investigated the crystal structure, magnetization, specific heat, and ESR studies of a quasi-two-dimensional frustrated magnet Y$_{2}$CuGe$_{4}$O$_{12}$, complemented by DFT calculations. The title compound crystallizes in a triclinic crystal structure without any detectable anti-site disorder, where the Cu$^{2+}$ ions form a distorted triangular lattice in the $ac$-plane.  Magnetic susceptibility measurements suggest predominant antiferromagnetic interactions between the spin-$\tfrac{1}{2}$ Cu$^{2+}$ moments. Magnetic susceptibility and specific heat measurements reveal no signature of long-range magnetic ordering  down to at least 0.4 K. Furthermore, the broad maxima observed in both magnetic susceptibility and specific heat indicate the development of short-range spin correlations, characteristic of a low-dimensional frustrated magnetic system, consistent with the ESR results. In addition, DFT calculations reveal spatially anisotropic exchange interactions consisting of both ferromagnetic and antiferromagnetic couplings over different Cu--Cu distances, resulting in an overall weak net magnetic energy scale consistent with the experimental observations. The weak net exchange interactions are further supported by the fact that a relatively small magnetic field of 2.6 T is sufficient to suppress the low-energy spin excitations and drive the system into a fully field-polarized state above the saturation field $\mu_{0}H_{\rm s}=2.6$ T  and the low-energy magnetic excitations develop a field-induced gap. Future low-temperature thermodynamic and spectroscopic  measurements on single crystals are required to reveal the magnetic ground state and to investigate  possible field induced phases, including magnetization plateaus.
 \\
 \begin{center}
 	\textbf{DATA AVAILABILITY}	
 \end{center}
The data supporting the findings of this study are
available upon reasonable request.
 \begin{center}
 	\textbf{ACKNOWLEDGMENTS}
 \end{center}
    The work at SKKU was supported by the National
    Research Foundation (NRF) of Korea (Grant No. RS2023-00209121, 2020R1A5A1016518). P.K. acknowledges the funding by the Anusandhan National Research Foundation (ANRF), Department of Science and
    Technology, India, through research grants.
\bibliography{YCGO.bib}
\end{document}